# Blowing Magnetic Skyrmion Bubbles

Wanjun Jiang[1], Pramey Upadhyaya[2], Wei Zhang[1], Guoqiang Yu[2], M. Benjamin Jungfleisch[1], Frank Y. Fradin[1], John E. Pearson[1], Yaroslav Tserkovnyak[3], Kang L. Wang[2], Olle Heinonen[1], Suzanne G. E. te Velthuis[1], and Axel Hoffmann[1,†]

[1]Materials Science Division, Argonne National Laboratory, Lemont, Illinois, USA, 60439

[2]Device Research Laboratory, Department of Electrical Engineering, University of California, Los Angeles, California, USA, 90095

[3]Department of Physics and Astronomy, University of California, Los Angeles, California, USA, 90095

[†]To whom correspondence should be addressed. E-mail: hoffmann@anl.gov.

**Soap bubbles form when blowing air through a suspended thin film of soapy water and this phenomenon entertains children and adults alike. The formation of soap bubbles from thin films is accompanied by topological transitions, and thus the natural question arises whether this concept is applicable to the generation of other topological states. Here we show how a magnetic topological structure, namely a skyrmion bubble, can be generated in a solid state system in a similar manner. Beyond enabling the investigation of complex surface-tension driven dynamics in a novel physical system, this observation has also practical implications, since the topological charge of magnetic skyrmions has been envisioned as an information carrier for new data processing technologies. A main goal**

**towards this end is the experimental creation and manipulation of individual mobile skyrmions at room temperature. By utilizing an inhomogeneous in-plane current in a system with broken inversion asymmetry, we experimentally "blow" magnetic skyrmion bubbles through a geometrical constriction. The presence of a spatially divergent spin-orbit torque gives rise to instabilities of the magnetic domain structures that are reminiscent of Rayleigh-Plateau instabilities in fluid flows. Experimentally we can determine the electric current versus magnetic field phase diagram for skyrmion formation and we reveal the efficient manipulation of these dynamically created skyrmions, including depinning and motion. The demonstrated current-driven transformation from stripe domains to magnetic skyrmion bubbles could provide additional avenues for implementing skyrmion-based spintronics.**

**One sentence summary:** Mobile magnetic skyrmions were experimentally created by spatially divergent in-plane electric currents at room temperature in a commonly accessible material system.

Magnetic skyrmions are topologically stable spin textures, which can be energetically favored by Dzyaloshinskii-Moriya interaction (DMI) (*1-9*) in chiral bulk magnets, e.g., MnSi, FeGe, *etc*. Due to their unique vortex-like spin-texture they exhibit many fascinating features including emergent electromagnetic fields, which enable their efficient manipulation (*4, 5, 8-10*). A particular technologically interesting fact is that skyrmions can be driven by a spin transfer torque mechanism at a very low current density, which has been demonstrated at cryogenic temperatures (*4, 5, 8, 10*). Besides bulk chiral magnetic interactions, the interfacial symmetry-breaking in heavy metal/ultra-thin ferromagnet/insulator (HM/F/I) trilayers with a perpendicular magnetic anisotropy introduces an interfacial DMI (*11-14*) between neighbouring atomic spins $\boldsymbol{S}_1$ and $\boldsymbol{S}_2$ that can be written as $-\boldsymbol{D}_{dmi} \cdot (\boldsymbol{S}_1 \times \boldsymbol{S}_2)$, where the DMI vector $\boldsymbol{D}_{dmi}$ lies in the film plane acting as an equivalent in-plane field and thus stabilizes Néel walls with a fixed chirality over the Bloch walls (*15-20*), as well as hedgehog skyrmions (*14, 18, 21-25*). This commonly accessible material system provides another distinct advantage by incorporating the spin Hall effects from heavy metals with strong spin-orbit interactions (*26*), which in turn give rise to well-defined spin-orbit torques (SOTs) (*17, 19, 27-29*) that can control magnetization dynamics efficiently. However, it has been experimentally challenging to utilize the electric current and/or its induced SOTs (*8, 21, 23, 24, 27, 30-32*) for dynamically creating and/or manipulating skyrmions. This missing ingredient, which is a pivotal step for realizing electrically programmable skyrmionic logic and/or memory at room temperature, is addressed in the present study.

Central to this work is how electric currents can manipulate a chiral magnetic domain wall (DW) that is stabilized by the interfacial DMI (*17-19, 21, 28*). In the HM/F/I heterostructures the current flowing through the heavy metal generates a transverse vertical spin current due to the spin Hall effect (*27*), which results in a spin accumulation at the interface with the ferromagnetic layer. This spin accumulation gives rise to a SOT acting on the chiral DW, as illustrated in Fig. 1(A). The resultant effective spin Hall field can be expressed as follows (*17-19, 27*):

$$\vec{\boldsymbol{B}}_{\text{sh}} = B_{\text{sh}}^0 \left( \widehat{m} \times (\hat{z} \times \hat{\jmath}_e) \right) \qquad (1)$$

where $\widehat{m}$ is the magnetization unit vector, $\hat{z}$ is the unit vector along the axis perpendicular to the film plane and $\hat{\jmath}_e$ is the direction of electron particle flux. Here $B_{\text{sh}}^0$ can be written as $(\hbar/2|e|) \cdot (\theta_{sh} J_c / t_f M_s)$, where $\hbar$ is the reduced Planck constant, $e$ is the electronic charge, $t_f$ is the thickness of the ferromagnetic layer, $M_s$ is the saturation (volume) magnetization. The spin Hall angle $\theta_{sh} = J_s/J_c$ is defined by the ratio between spin current density ($J_s$) and charge current density ($J_c$). Given a homogeneous current flow along $x$ axis, as depicted in Fig. 1 (B), a chiral SOT enables efficient DW motion (*17-19*). In the case of a stripe domain with a chiral DW, the symmetry of Eq. (1) leads to a vanishing torque on the side walls parallel to the current and therefore only the end of the stripe domain is moved resulting in an elongation of the stripe, if the opposite end is pinned.

However, the situation becomes more complex, when the stripe domain is subjected to an inhomogeneous current flow. We address this scenario by introducing a geometrical constriction into a current carrying trilayer wire, as shown in Fig. 1(C). Such a geometrical constriction results in an additional (spatially convergent/divergent) current components along the $y$ axis - $j_y$ around the narrow neck, as shown in Fig. 1(D). This is furthermore confirmed via a finite element simulation by using the CST Electro-Magnetic Studio software, as discussed in the supplementary information. Consequently, inhomogeneous effective forces on the DWs (due to the spin Hall field), are created along the $y$ axis - $\boldsymbol{F}_{sh}^y$, which expand the domain, as shown in Fig. 1(E). As the domain continually expands its radius, due to the outward "pressure" from the effective spin Hall field, the surface tension in the DW (resulting from the increasing DW energy determined by the combination of exchange and anisotropy fields) increases (*33*), which results in breaking the stripes into circular domains, Fig. 1(F).

This process resembles how soap bubbles develop out of soap films upon blowing air through a straw, or how liquid droplets form in fluid flow jets (*34*). Due to the interfacial DMI in the present system, the spin structures of the newly formed circular domains therefore maintain a well-defined (left-handed) chirality (*13, 14, 23, 24*). Therefore, these dynamically created (synthesized) hedgehog/Néel skyrmions (*14, 23*), once formed, are stable due to topological protection and move very efficiently following the current direction that can be formulated based on a modified Thiele equation (*35*). The dynamic skyrmion conversion could, in principle, happen at the other side of device where the spatially convergent current compresses stripe domains. However, significant currents/SOTs are required to compensate the enhanced (repulsive) dipolar interaction. Also note that this mechanism differs from a recent theoretical proposal with similar geometry, where skyrmions are formed from the coalescence of two independent DWs extending over the full width of a narrow constriction at a current density ≈ $10^8$ A/cm$^2$ (*32*). For repeated skyrmion generation, this latter mechanism requires a continuous generation of paired DWs in the constriction, which is inconsistent with the experimental observations described below.

**Transforming chiral stripe domains into skyrmions.**

In order to demonstrate experimentally this idea we used a Ta(5nm)/Co$_{20}$Fe$_{60}$B$_{20}$(CoFeB)(1.1nm)/TaO$_x$(3nm) trilayer grown by magnetron sputtering (*36, 37*) and patterned into constricted wires via photolithography and ion-milling. The wires have width 60 μm with a geometrical constriction of width 3 μm and of length 20 μm in the center. Our devices are symmetrically designed across the narrow neck to maintain balanced demagnetization energy. A polar magneto-optical Kerr effect (MOKE) microscope in a differential mode (*38*) was utilized for dynamic imaging experiments at room temperature. Before applying a current, the sample was first saturated at positive magnetic fields and subsequently at a perpendicular magnetic field of $B_\perp = +0.5$ mT, sparse magnetic stripe and bubble domains prevail at both sides of the wire, as shown in Fig. 2(A). The lighter area corresponds to negative perpendicular magnetization orientation and darker area corresponds to positive orientation, respectively.

In contrast to the initial magnetic domain configuration, after passing a 1 s single pulse of amplitude $j_e = +5 \times 10^5$ A/cm$^2$ (normalized by the width of device – 60 µm), it is observed that the stripe domains started to migrate, subsequently forming extended stripe domains on the left side, which mostly aligned with the charge current flow and converged at the left side of constriction. Interestingly at the right side of the device the stripes are transformed into skyrmion bubbles immediately after passing through the constriction, as shown in Fig. 2(B). These dynamically created skyrmions, varying in size between 700 nm and 2 µm (depending on the strength of the external field), once formed, are stable and do not decay for the typical laboratory testing period (8 hours at least). It is also noted that in the presence of a constant electron current density of $j_e = +5 \times 10^5$ A/cm$^2$, these skyrmions are created close to the central constriction and annihilated/destroyed at the end of the wire with a high speed. Capturing the transformation dynamics of skyrmions from stripe domains is beyond the temporal resolution of the present setup. In addition, reproducible generation of skyrmions is also demonstrated by repeating pulsed experiments for several times, selective results of which are presented in the Supplementary Materials. Furthermore, it is also interesting to mention that the left side of device remains mainly in the labyrinthine stripe domain state after removing the pulse current, which indicates that both skyrmion bubbles and stripe domains are allowed in equilibrium.

By simply reversing the polarity of the charge current to $j_e = -5\times10^5$ A/cm$^2$, these skyrmions are formed at the left side of device as opposed to the positive current, see Figs. 2(C)–(D). This directional dependence feature indicates that the spatially divergent current/SOT, determined by the geometry of the device, is most likely responsible for slicing stripe DWs into magnetic skyrmion bubbles, qualitatively consistent with the schematic presented in Fig. 1.

Selected experiments at a negative magnetic field $B_\perp = -0.5$ mT and current at $j_e = +5\times10^5$ A/cm$^2$, are given in Figs. 2(E)–(F), in which the reversal of inner/outer magnetization orientations are observed as compared with positive fields. Systematic studies on the evolution of skyrmion formation as a function of the external magnetic fields and charge currents/SOTs are carried out which consequently establishes a phase diagram, as shown in Fig. 2(G). A large population of synthetic skyrmions is found only in the shadowed region, while in the rest of phase diagram, the initial domain configurations remain either stationary or flowing smoothly, depending the strength of current density, as discussed below. This phase diagram is independent of pulse duration for pulses longer than 1 μs. It should be mentioned that no creation of skyrmions in regular shaped device with a homogeneous current flow (as illustrated in Fig. 1(B)) is observed up to a current density $j_e = +5\times10^6$ A/cm$^2$.

**Capturing the transformational dynamics**

An experimental visualization of the conversion from chiral stripe domains into magnetic skyrmions can be captured by decreasing the driving current. This slows down the transformational dynamics. Figures 3(A)–(D) show this for a constant *dc* current density of $j_e$ = +6.4×10$^4$ A/cm$^2$ at $B_\perp$ = +0.46 mT. The original (disordered) labyrinthine domains, at the left side are converged/squeezed to pass through the constriction, see Fig. 3(B). Due to the divergent SOTs, the stripe domains become unstable after passing through the constriction and are eventually converted into skyrmions at the right side of the device, as shown in Figs. 3(C)–(D). This can be seen in more detail in the MOKE Movies (S1 and S2) given in the Supplementary Materials. Due to the *x*-component of current which results in the efficient motion of DWs, the skyrmion formation can happen somewhat displaced away from the constriction. It is noted that these synthetic skyrmions do not merge into stripe domains and in fact repel each other, indicating their topological protection as well as magnetostatic interactions.

Some important features should be noticed. There exists a threshold current $j_{e-sk}$ = ±6×10$^4$ A/cm$^2$ for persistently generating skyrmion bubbles from stripe domains for each pulse down to 1 μs. Above this current the enhanced spin-orbit torques produce the instability of the DWs, which results in the continuous formation of skyrmions. It should be highlighted here that the present geometry for skyrmion generation is very efficient resulting in the observed threshold current of 3 orders of magnitude smaller than suggested by previous simulation studies (10$^{7-8}$ A/cm$^2$) in MnSi thin films with a bulk DMI where the driven mechanism is the conventional spin transfer torque (*30*). Below this threshold for continuous skyrmion generation, there is a threshold depinning current $j_{e-st}$ = ±4.1×10$^4$ A/cm$^2$ that produces a steady motion of stripe domains. The force (pressure) on the stripe from SOT at this current exceeds the one required to maintain its shape. When $j_{e-st} < j_e < j_{e-sk}$, the stripe domains are moving smoothly through the constriction and prevail at both sides of devices, with just the occasional formation of skyrmions.

**Depinning and Motion of synthesized $S = 1$ skyrmions**

The magnetic skyrmion bubbles discussed so far have a topological charge given by the skyrmion number $S = 1$, as is determined by wrapping the unit magnetization vector over the sphere $S = \frac{1}{4\pi} \int \boldsymbol{m} \cdot (\partial_x \boldsymbol{m} \times \partial_y \boldsymbol{m}) \, dxdy$ (*1, 8*). These $S = 1$ synthetic skyrmions move due to the opposite direction of effective SOTs at opposite sides of the skyrmion, as schematically illustrated in Fig. 3(E). Following the initialization by a passing current $j_e = +5 \times 10^5$ A/cm$^2$ (which is larger than the threshold current $j_{e-sk}$ for generating skyrmions), the efficient depinning and motion of synthetic skyrmions are subsequently studied and given in Figs. 3(F)–(I) at $B_\perp = -0.5$ mT. At the current density $j_e = +3 \times 10^4$ A/cm$^2$, there is no migration of stripe domain through the constriction (hence an absence of newly-formed skyrmions). It is however, clear to see that the previously generated skyrmions at the right side of the device are gradually moving away following the electron flow direction. During the motion, no measurable distortion of these synthetic skyrmions is observed within the experimental resolution, consistent with the well-defined chirality of the skyrmion bubble. The average velocity ($\bar{v} = \ell/\Delta t$) is determined by dividing the displacement ($\ell$) with the total time period ($\Delta t$). For the present current density, the motion of synthetic skyrmion is stochastic and influenced by random pinning with an average velocity of about 10 µm/s. The current dependence of which is summarized in Fig. 3(J). It should be noted that the ratio of velocity to applied current is comparable to what is observed for the chiral DW motion in the related systems (*17, 19*).

**Current characteristics of $S = 0$ magnetic bubbles**

Due to the competition between long-range dipolar and short-range exchange interaction, a system with weak perpendicular magnetic anisotropy undergoes a spin reorientation transition with in-plane magnetic fields that is typified by a stripe-to-bubble domain phase transition (*38, 39*). Such an in-plane field induced bubble state is established by sweeping magnetic field from $B_\parallel = +100$ mT to $B_\parallel = +10$ mT. In contrast to the mobile magnetic skyrmions generated from SOTs, current-driven characteristics of the in-plane field induced magnetic bubbles are intriguing. Namely, these bubbles shrink and vanish in the presence of a positive electron current density, see Figs. 4(A)–(E), or elongate and transform into stripe domains in the presence of negative electron current density, see Figs. 4(F)–(J). Such a distinct difference directly indicates the different spin structures surrounding these field induced bubbles, and thereby different skyrmion numbers.

For the in-plane field induced magnetic bubbles, since the spin structures of DWs follow the external magnetic fields (*18, 40, 41*), as schematically illustrated in Fig. 4(K), the corresponding skyrmion number is determined to be $S = 0$. Due to the same direction of the spin Hall effective fields given by the reversed DW orientations, topologically trivial $S = 0$ magnetic bubbles therefore experience opposite forces on the DWs at opposite ends. This leads to either a shrinking or elongation of the bubbles depending on the direction of currents, which is consistent with our experimental observation. It should be noted here that this also explains well the in-plane current induced perpendicular magnetization switching in the presence of in-plane fields (*27, 41*).

**Perspectives**

Recent experimental efforts towards creating individual magnetic skyrmions use either tunnelling current from a low-temperature spin-polarized scanning tunnelling microscope (*42*) or geometrical confinement via sophisticated nanopatterning (*43-45*). Our result, unambiguously demonstrate that spatially divergent current-induced SOTs can be an effective way for dynamically generating mobile magnetic skyrmions at room temperature in commonly accessible material systems, which therefore, constitutes an important step towards skyrmion-based spintronics – skyrmionics. It should mention here that the size of these synthetic skyrmions could be scaled down by properly engineering the material specific parameters that control the various competing interactions in magnetic nanostructures (*23, 24, 46*). We expect that similar instabilities will be generated from divergent charge current flows. While the mechanism for synthetic skyrmion generation can be qualitatively linked to the spatially divergent spin Hall spin torque, a comprehensive understanding of this dynamical conversion, particularly at the picosecond/nanosecond time scale where the intriguing magnetization dynamics occurs, requires further experimental and theoretical investigations. Therefore, spatially divergent SOT-driven structures also offer a readily accessible model system for studying topological transitions and complex "flow" instabilities (*34*), where the parameters governing the flow, such as surface tension, can be systematically tuned by the magnetic interactions. At the same time, this dynamic approach for skyrmion generation in the near future could enable the demonstration of advanced skyrmionic device concepts, for example, functional skyrmion racetrack memory (*14, 23, 35, 47*).

**References and notes**

1. U. K. Rossler, A. N. Bogdanov, C. Pfleiderer, Spontaneous skyrmion ground states in magnetic metals. *Nature* **442**, 797-801 (2006).
2. S. Muhlbauer, B. Binz, F. Jonietz, C. Pfleiderer, A. Rosch, A. Neubauer, R. Georgii, P. Boni, Skyrmion Lattice in a Chiral Magnet. *Science* **323**, 915-919 (2009).
3. X. Z. Yu, Y. Onose, N. Kanazawa, J. H. Park, J. H. Han, Y. Matsui, N. Nagaosa, Y. Tokura, Real-space observation of a two-dimensional skyrmion crystal. *Nature* **465**, 901-904 (2010).
4. F. Jonietz, S. Muhlbauer, C. Pfleiderer, A. Neubauer, W. Munzer, A. Bauer, T. Adams, R. Georgii, P. Boni, R. A. Duine, K. Everschor, M. Garst, A. Rosch, Spin Transfer Torques in MnSi at Ultralow Current Densities. *Science* **330**, 1648-1651 (2010).
5. X. Z. Yu, N. Kanazawa, Y. Onose, K. Kimoto, W. Z. Zhang, S. Ishiwata, Y. Matsui, Y. Tokura, Near room-temperature formation of a skyrmion crystal in thin-films of the helimagnet FeGe. *Nature Materials* **10**, 106-109 (2011).
6. S. Seki, X. Z. Yu, S. Ishiwata, Y. Tokura, Observation of Skyrmions in a Multiferroic Material. *Science* **336**, 198-201 (2012).
7. H. B. Braun, Topological effects in nanomagnetism: from superparamagnetism to chiral quantum solitons. *Advances in Physics* **61**, 1-116 (2012).
8. N. Nagaosa, Y. Tokura, Topological properties and dynamics of magnetic skyrmions. *Nature Nanotechnology* **8**, 899-911 (2013).
9. S. Z. Lin, C. Reichhardt, C. D. Batista, A. Saxena, Driven Skyrmions and Dynamical Transitions in Chiral Magnets. *Physical Review Letters* **110**, 207202 (2013).
10. J. D. Zang, M. Mostovoy, J. H. Han, N. Nagaosa, Dynamics of Skyrmion Crystals in Metallic Thin Films. *Physical Review Letters* **107**, 136804 (2011).
11. M. Bode, M. Heide, K. von Bergmann, P. Ferriani, S. Heinze, G. Bihlmayer, A. Kubetzka, O. Pietzsch, S. Blugel, R. Wiesendanger, Chiral magnetic order at surfaces driven by inversion asymmetry. *Nature* **447**, 190-193 (2007).
12. S. Heinze, K. von Bergmann, M. Menzel, J. Brede, A. Kubetzka, R. Wiesendanger, G. Bihlmayer, S. Blugel, Spontaneous atomic-scale magnetic skyrmion lattice in two dimensions. *Nature Physics* **7**, 713-718 (2011).
13. A. Thiaville, S. Rohart, E. Jue, V. Cros, A. Fert, Dynamics of Dzyaloshinskii domain walls in ultrathin magnetic films. *Epl* **100**, 57002 (2012).
14. A. Fert, V. Cros, J. Sampaio, Skyrmions on the track. *Nature Nanotechnology* **8**, 152-156 (2013).
15. G. Chen, T. P. Ma, A. T. N'Diaye, H. Kwon, C. Won, Y. Z. Wu, A. K. Schmid, Tailoring the chirality of magnetic domain walls by interface engineering. *Nature Communications* **4**, 3671 (2013).
16. G. Chen, J. Zhu, A. Quesada, J. Li, A. T. N'Diaye, Y. Huo, T. P. Ma, Y. Chen, H. Y. Kwon, C. Won, Z. Q. Qiu, A. K. Schmid, Y. Z. Wu, Novel Chiral Magnetic Domain Wall Structure in Fe/Ni/Cu(001) Films. *Physical Review Letters* **110**, 177204 (2013).
17. S. Emori, U. Bauer, S. M. Ahn, E. Martinez, G. S. Beach, Current-driven dynamics of chiral ferromagnetic domain walls. *Nature Materials* **12**, 611-616 (2013).
18. S. Emori, E. Martinez, K.-J. Lee, H.-W. Lee, U. Bauer, S.-M. Ahn, P. Agrawal, D. C. Bono, G. S. D. Beach, Spin Hall torque magnetometry of Dzyaloshinskii domain walls. *Physical Review B* **90**, 184427 (2014).

19. K. S. Ryu, L. Thomas, S. H. Yang, S. Parkin, Chiral spin torque at magnetic domain walls. *Nature Nanotechnology* **8**, 527-533 (2013).
20. O. Boulle, S. Rohart, L. D. Buda-Prejbeanu, E. Jue, I. M. Miron, S. Pizzini, J. Vogel, G. Gaudin, A. Thiaville, Domain Wall Tilting in the Presence of the Dzyaloshinskii-Moriya Interaction in Out-of-Plane Magnetized Magnetic Nanotracks. *Physical Review Letters* **111**, 217203 (2013).
21. N. Perez, E. Martinez, L. Torres, S. H. Woo, S. Emori, G. S. D. Beach, Chiral magnetization textures stabilized by the Dzyaloshinskii-Moriya interaction during spin-orbit torque switching. *Applied Physics Letters* **104**, 092403 (2014).
22. K. W. Kim, H. W. Lee, K. J. Lee, M. D. Stiles, Chirality from Interfacial Spin-Orbit Coupling Effects in Magnetic Bilayers. *Physical Review Letters* **111**, 216601 (2013).
23. J. Sampaio, V. Cros, S. Rohart, A. Thiaville, A. Fert, Nucleation, stability and current-induced motion of isolated magnetic skyrmions in nanostructures. *Nature Nanotechnology* **8**, 839-844 (2013).
24. S. Rohart, A. Thiaville, Skyrmion confinement in ultrathin film nanostructures in the presence of Dzyaloshinskii-Moriya interaction. *Physical Review B* **88**, 184422 (2013).
25. B. Dupe, M. Hoffmann, C. Paillard, S. Heinze, Tailoring magnetic skyrmions in ultra-thin transition metal films. *Nature Communications* **5**, 4030 (2014).
26. A. Hoffmann, Spin Hall Effects in Metals. *IEEE Transactions on Magnetics* **49**, 5172-5193 (2013).
27. L. Q. Liu, C. F. Pai, Y. Li, H. W. Tseng, D. C. Ralph, R. A. Buhrman, Spin-Torque Switching with the Giant Spin Hall Effect of Tantalum. *Science* **336**, 555-558 (2012).
28. A. V. Khvalkovskiy, V. Cros, D. Apalkov, V. Nikitin, M. Krounbi, K. A. Zvezdin, A. Anane, J. Grollier, A. Fert, Matching domain-wall configuration and spin-orbit torques for efficient domain-wall motion. *Physical Review B* **87**, 020402 (2013).
29. I. M. Miron, K. Garello, G. Gaudin, P. J. Zermatten, M. V. Costache, S. Auffret, S. Bandiera, B. Rodmacq, A. Schuhl, P. Gambardella, Perpendicular switching of a single ferromagnetic layer induced by in-plane current injection. *Nature* **476**, 189-193 (2011).
30. J. Iwasaki, M. Mochizuki, N. Nagaosa, Current-induced skyrmion dynamics in constricted geometries. *Nature Nanotechnology* **8**, 742-747 (2013).
31. Y. Tchoe, J. H. Han, Skyrmion generation by current. *Physical Review B* **85**, 174416 (2012).
32. Y. Zhou, M. Ezawa, A reversible conversion between a skyrmion and a domain-wall pair in a junction geometry. *Nature Communications* **5**, 4652 (2014).
33. A. P. Malozemoff, J. C. Slonczewski, *Magnetic Domain Walls in Bubble Materials*. (Academic Press, New York, 1979).
34. J. Eggers, Nonlinear dynamics and breakup of free-surface flows. *Reviews of Modern Physics* **69**, 865-930 (1997).
35. R. Tomasello, E. Martinez, R. Zivieri, L. Torres, M. Carpentieri, G. Finocchio, A strategy for the design of skyrmion racetrack memories. *Scientific Reports* **4**, 6784 (2014).
36. G. Q. Yu, P. Upadhyaya, Y. B. Fan, J. G. Alzate, W. J. Jiang, K. L. Wong, S. Takei, S. A. Bender, L. T. Chang, Y. Jiang, M. R. Lang, J. S. Tang, Y. Wang, Y. Tserkovnyak, P. K. Amiri, K. L. Wang, Switching of perpendicular magnetization by spin-orbit torques in the absence of external magnetic fields. *Nature Nanotechnology* **9**, 548-554 (2014).

37. G. Q. Yu, P. Upadhyaya, K. L. Wong, W. J. Jiang, J. G. Alzate, J. S. Tang, P. K. Amiri, K. L. Wang, Magnetization switching through spin-Hall-effect-induced chiral domain wall propagation. *Physical Review B* **89**, 104421 (2014).
38. A. Hubert, R. Schafer, *Magnetic Domains: The Analysis of Magnetic Microstructures* (Springer, Berlin, Heidelberg, New York, 2008).
39. J. Choi, J. Wu, C. Won, Y. Z. Wu, A. Scholl, A. Doran, T. Owens, Z. Q. Qiu, Magnetic bubble domain phase at the spin reorientation transition of ultrathin Fe/Ni/Cu(001) film. *Physical Review Letters* **98**, 207205 (2007).
40. J. H. Franken, M. Herps, H. J. M. Swagten, B. Koopmans, Tunable chiral spin texture in magnetic domain-walls. *Scientific Reports* **4**, 5248 (2014).
41. O. J. Lee, L. Q. Liu, C. F. Pai, Y. Li, H. W. Tseng, P. G. Gowtham, J. P. Park, D. C. Ralph, R. A. Buhrman, Central role of domain wall depinning for perpendicular magnetization switching driven by spin torque from the spin Hall effect. *Physical Review B* **89**, 024418 (2014).
42. N. Romming, C. Hanneken, M. Menzel, J. E. Bickel, B. Wolter, K. von Bergmann, A. Kubetzka, R. Wiesendanger, Writing and Deleting Single Magnetic Skyrmions. *Science* **341**, 636-639 (2013).
43. L. Sun, R. X. Cao, B. F. Miao, Z. Feng, B. You, D. Wu, W. Zhang, A. Hu, H. F. Ding, Creating an Artificial Two-Dimensional Skyrmion Crystal by Nanopatterning. *Physical Review Letters* **110**, 167201 (2013).
44. J. Li, A. Tan, K. W. Moon, A. Doran, M. A. Marcus, A. T. Young, E. Arenholz, S. Ma, R. F. Yang, C. Hwang, Z. Q. Qiu, Tailoring the topology of an artificial magnetic skyrmion. *Nature Communications* **5**, 4704 (2014).
45. B. F. Miao, L. Sun, Y. W. Wu, X. D. Tao, X. Xiong, Y. Wen, R. X. Cao, P. Wang, D. Wu, Q. F. Zhan, B. You, J. Du, R. W. Li, H. F. Ding, Experimental realization of two-dimensional artificial skyrmion crystals at room temperature. *Physical Review B* **90**, 174411 (2014).
46. A. Hrabec, N. A. Porter, A. Wells, M. J. Benitez, G. Burnell, S. McVitie, D. McGrouther, T. A. Moore, C. H. Marrows, Measuring and tailoring the Dzyaloshinskii-Moriya interaction in perpendicularly magnetized thin films. *Physical Review B* **90**, 020402(R) (2014).
47. S. S. P. Parkin, M. Hayashi, L. Thomas, Magnetic domain-wall racetrack memory. *Science* **320**, 190-194 (2008).

**Acknowledgements:**

Work carried out at the Argonne National Laboratory was supported by the U.S. Department of Energy, Office of Science, Materials Science and Engineering Division. Lithography was carried out at the Center for Nanoscale Materials, which is supported by the DOE, Office of Science, Basic Energy Science under Contract No. DE-AC02-06CH11357. Work performed at UCLA was partially supported by the NSF Nanosystems Engineering Research Center for Translational

Applications of Nanoscale Multiferroic Systems (TANMS). The authors also acknowledge insightful discussion with Ivar Martin and Igor Aronson.

**Figure captions:**

**Figure 1. Schematic of the transformation of stripe domains into magnetic skyrmion bubbles.** (A) Infinitesimal section of a chiral DW illustrating the relationship between local magnetization vectors and the SOT-induced chiral DW motion of velocity $V_{dw}$ in a device with a homogeneous electron current flow $j_e$ along the *x* axis. (B) The left-handed chirality due to the presence of interfacial DMI is depicted by the light blue arrows indicating the magnetization direction at the DW center. (C) Introducing a geometrical constriction into the devices gives rise to an inhomogeneous current distribution, which generates a flow along the *y*-axis, $j_y$ around the narrow neck. This current distribution is spatially divergent to the right and convergent to the left of the constriction. The *y*-component of the current distribution is highlighted in (D). This introduces an effective spin Hall force $\boldsymbol{F}_{sh}^{y}$ along the *y*-axis that (E) locally expands the stripe domains on the right side. (F) Once the expansion approaches a critical point, the resultant restoring forces $\boldsymbol{F}_{res}$ that are associated with the surface tension of the DWs, are no longer able to maintain the shape and the stripe domains break into circular bubble domains, resulting in the formation of synthesized Néel skyrmions.

**Figure 2. Experimental generation of magnetic skyrmions.** (A) Sparse irregular domain structures are observed at both sides of the device at a perpendicular magnetic field of $B_\perp = +0.5$ mT. (B) Upon passing a current of $j_e = +5 \times 10^5$ A/cm$^2$ through the device, the left side of the device develops predominantly elongated stripe domains, while the right side converts into dense

skyrmion bubbles. (C)–(D) By reversing the current direction to $j_e$ = -5×10$^5$ A/cm$^2$, the dynamically created skyrmions are forming at the left side of device. (E)–(F) In addition, changing the polarity of external magnetic field also reverses the internal and external magnetization of these skyrmions. (G) Systematic studies at different fields and threshold current amplitudes generate a phase diagram for skyrmion formation. The shaded area indicates field/current combinations that result in the persistent generation of skyrmions after each current pulse.

**Figure 3. Capturing the transformational dynamics from stripe domains to skyrmions and motion of skyrmions.** (A)–(D) At a constant *dc* current $j_e$ = +6.4×10$^4$ A/cm$^2$ and $B_\perp$ = +0.46 mT, the disordered stripe domains are forced to pass through the constriction, and are eventually converted into skyrmions at the right side of the device. The corresponding temporal dynamics is provided in the Supplementary Materials. Red circles highlight the resultant newly-formed skyrmions. (E) Schematic illustration of the spin-Hall effective field acting on these dynamically created skyrmions; namely, the direction of motion follows the electron current. (F)–(I) The efficient motion of these skyrmions for a current density $j_e$ = +3×10$^4$ A/cm$^2$. (F) First, a 1 s long single pulse $j_e$ = +5×10$^5$ A/cm$^2$ initializes the skyrmion state. (G)–(I) Subsequently, smaller currents (below the threshold current to avoid generating additional skyrmions through the constriction) are used to probe the current-velocity relation. It is observed that these skyrmions are migrating stochastically, and skyrmions are moving out of the field of view. (J) The current-velocity dependence of skyrmions is acquired by studying approximately 20 skyrmions via averaging their velocities by dividing the total displacement with the total time period.

**Figure 4. Absence of motion for the in-plane magnetic fields stabilized $S = 0$ magnetic bubbles.** (A) In-plane magnetic field induced bubbles are created by first saturating at in-plane field $B_\parallel = +100$ mT and subsequently decreasing to $B_\parallel = +10$ mT. Depending on the direction of the current, these magnetic bubbles either shrink or expand. (A) – (E) The shrinking bubbles are observed upon increasing the current density from $j_e = +5\times10^4$ A/cm$^2$ to $+2.5\times10^5$ A/cm$^2$ in $5\times10^4$ A/cm$^2$ steps. (F)–(J) The expansion of bubbles is revealed for currents from $j_e = -0.5\times10^5$ A/cm$^2$ to $-2.5\times10^5$ A/cm$^2$ in $5\times10^4$ A/cm$^2$ steps. (K) These results are linked to the different spin textures that were stabilized along the DW by the in-plane magnetic fields, namely, $S = 0$ skyrmion bubbles, which lead to different orientations of the spin Hall effective fields and different directions of DW motion as illustrated.

**Supplementary Materials:**

Figures S1-S7

Movies S1-S5

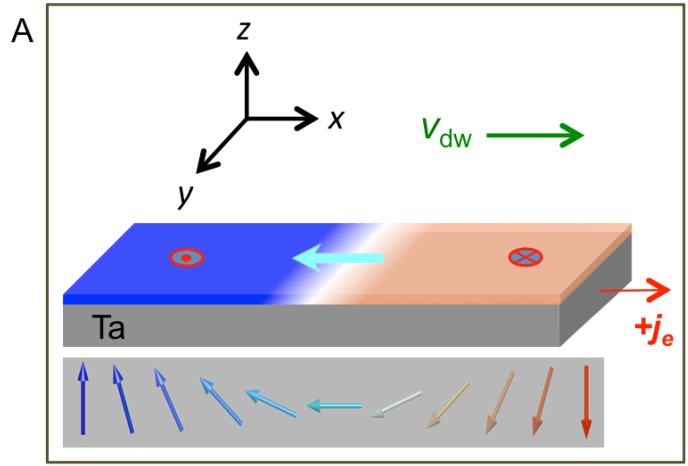
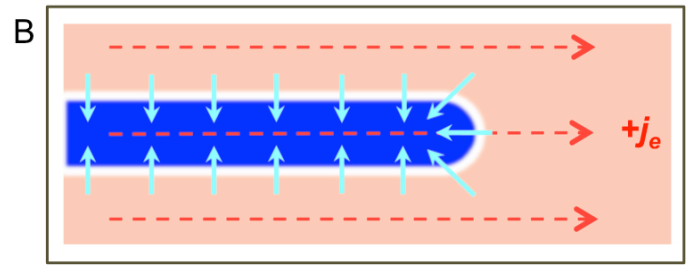
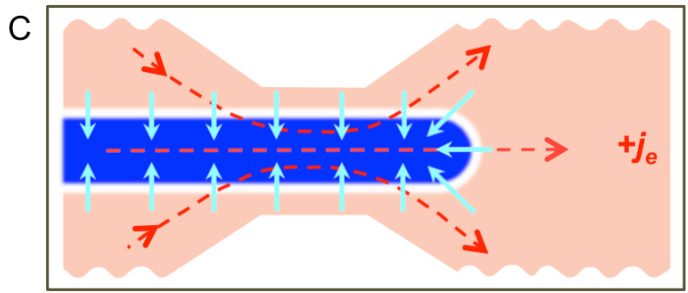
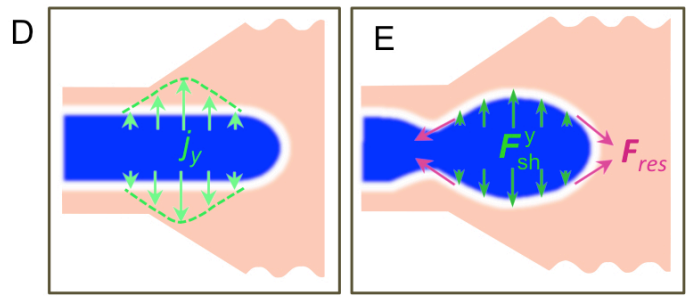
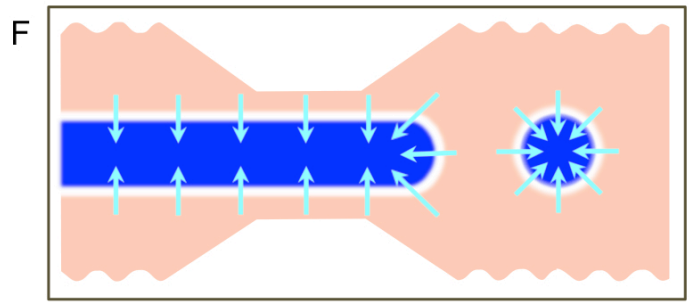

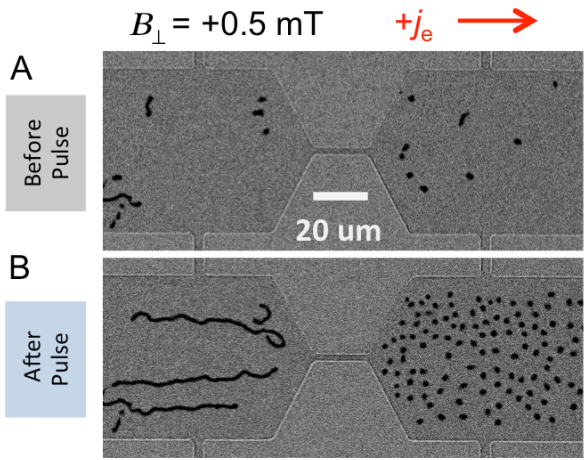
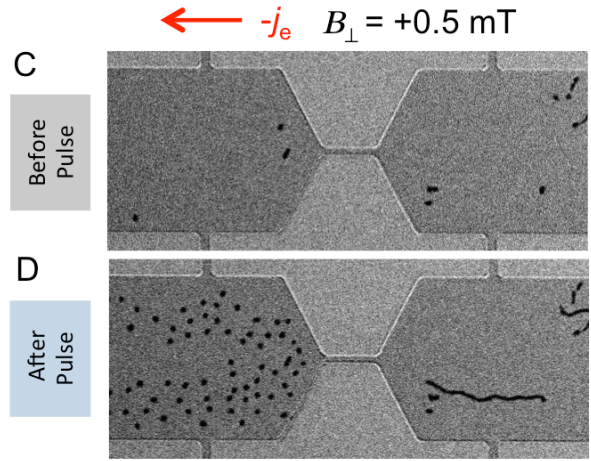
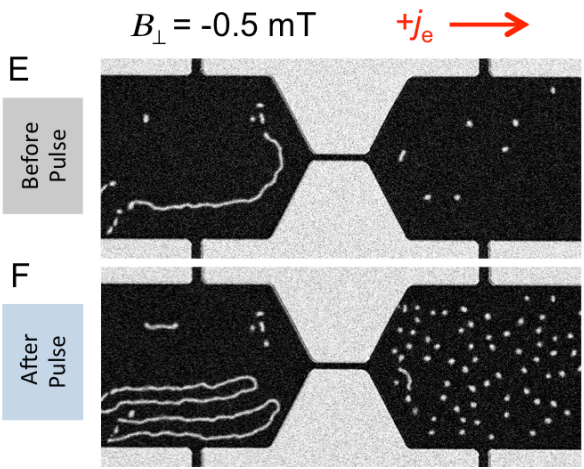
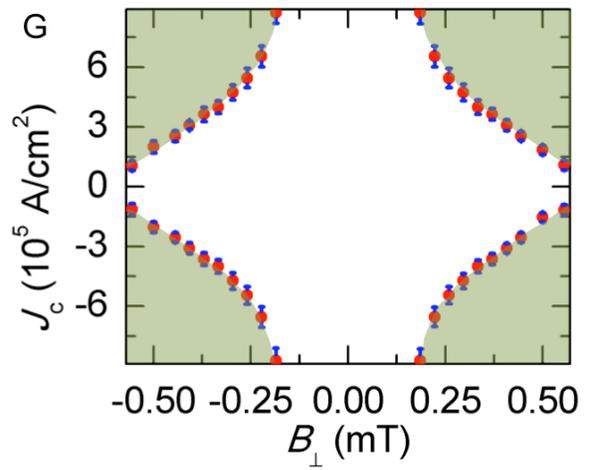

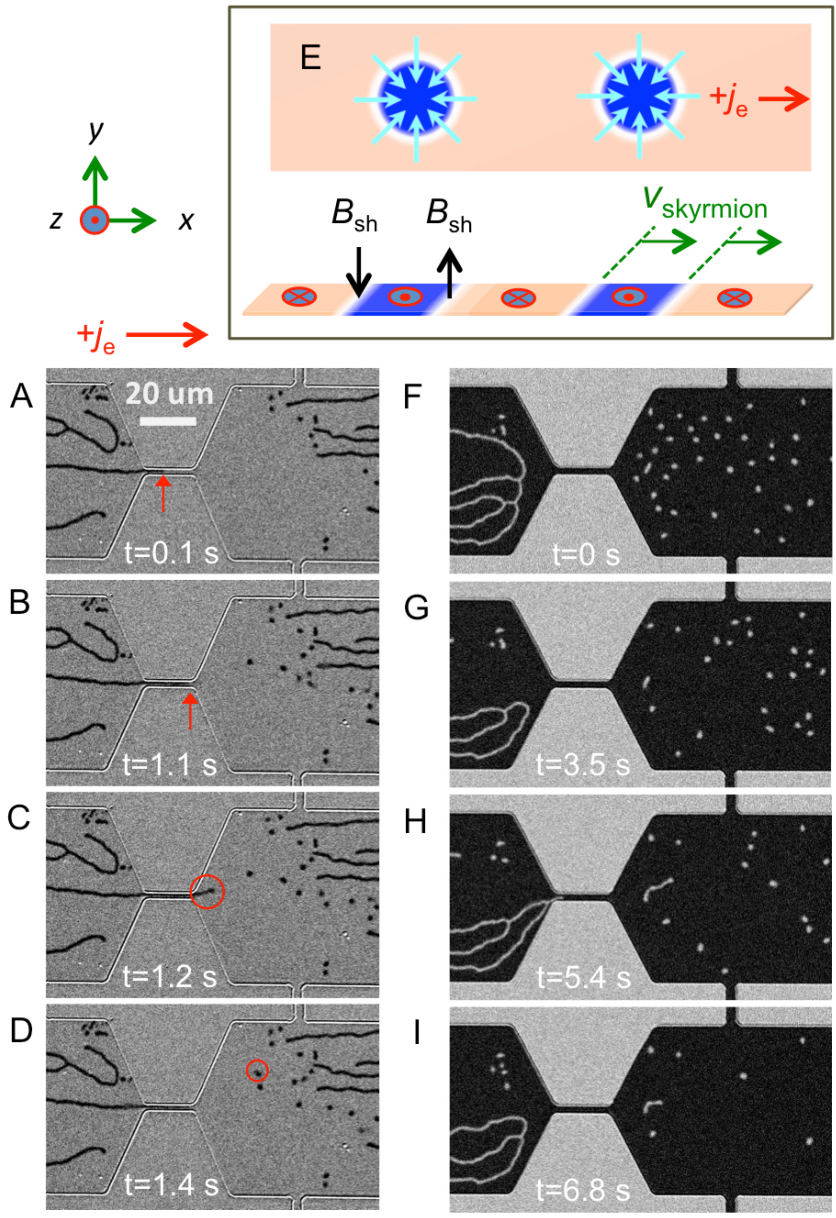
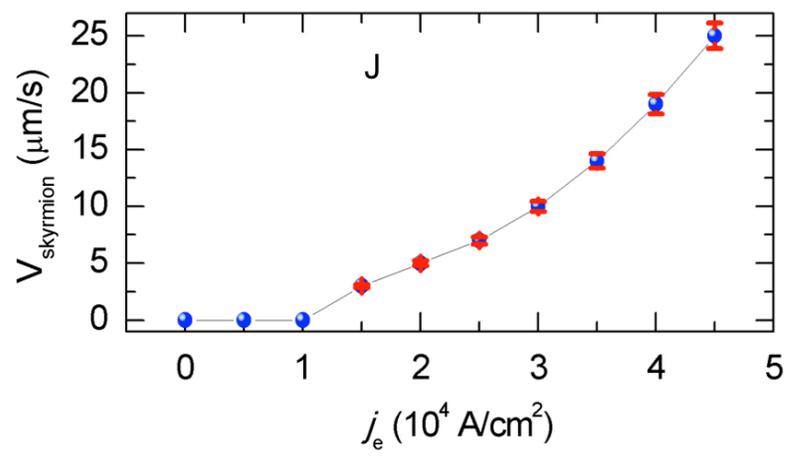

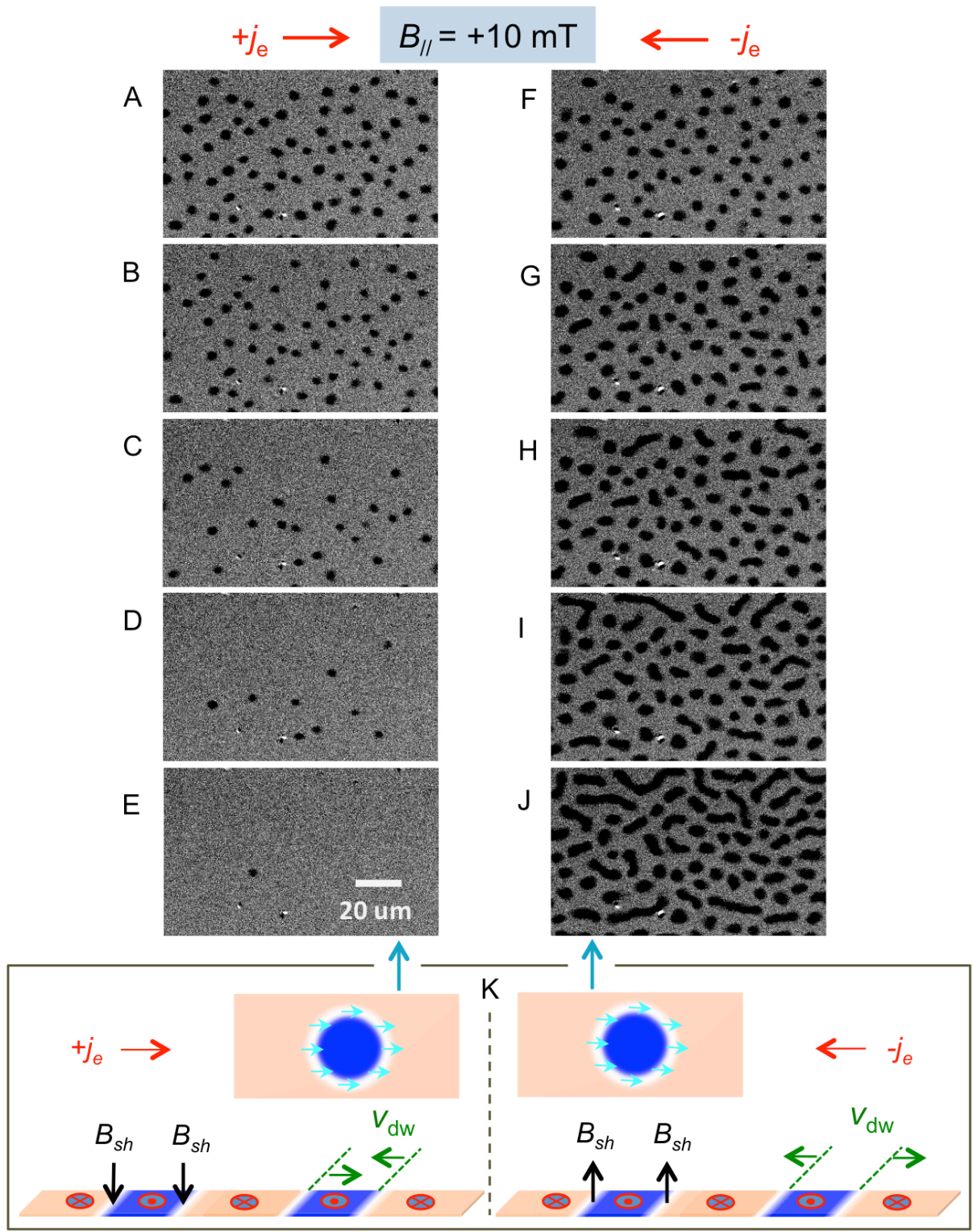